\documentclass[aps,prl,onecolumn,showpacs,preprintnumbers]{revtex4}
\usepackage{eurosym}
\usepackage{amsmath}
\usepackage{dcolumn}
\usepackage{bm}
\usepackage{subfigure}
\usepackage{amsfonts}
\usepackage{amssymb}
\usepackage{makeidx}
\usepackage{epsfig}
\usepackage{graphicx}

\begin{document}

\title{\textbf{Principles of kinetic theory for granular fluids\footnote{%
Based on a seminar presented at the Weizmann Institute of Science (Rehovot,
Israel), May 4, 2015}}}
\author{Massimo Tessarotto}
\affiliation{Department of Mathematics and Geosciences, University of Trieste, Via
Valerio 12, 34127 Trieste, Italy}
\affiliation{Institute of Physics, Faculty of Philosophy and Science, Silesian University
in Opava, Bezru\v{c}ovo n\'{a}m.13, CZ-74601 Opava, Czech Republic}
\author{Claudio Cremaschini}
\affiliation{Institute of Physics, Faculty of Philosophy and Science, Silesian University
in Opava, Bezru\v{c}ovo n\'{a}m.13, CZ-74601 Opava, Czech Republic}
\date{\today }

\begin{abstract}
Highlights are presented regarding recent developments of the kinetic theory
of granular matter. These concern the discovery of an exact kinetic equation
and a related exact H-theorem both holding for finite $N-$body systems
formed by smooth hard-spheres systems.
\end{abstract}

\maketitle

%\tableofcontents

%\frontmatter

\section{Introduction}

Although Ludwig Boltzmann's discovery of his namesake equation and H-theorem
dates back to more than a century \cite{Boltzmann1972}, only a few papers
have actually given a significant contribution to the advancement of kinetic
theory itself. This refers in particular to the treatment of dense or
granular systems, \textit{i.e.,} in which the finite size of constituent
particles must be taken into account, starting from the phenomenological
Enskog kinetic equation originally formulated by Enskog \cite{Enskog} for
elastic smooth hard-spheres and its subsequent modified form \cite{van
Beijeren}. In fact, in spite of the progress achieved in describing the
kinetics and hydrodynamics of granular fluids, represented in particular by
non-linear theories such as the Bogoliubov-Choh-Uhlenbeck theory \cite{Cohen}
and the so-called ring kinetic theory \cite{Dorfman}, the solution of this
problem has not changed significantly and has remained until recently "\emph{%
far from being complete}" \cite{Goldhirsh2000}.

Concerning, instead, the treatment of dilute gases an exception is provided
by Harold Grad's seminal paper on the \textit{Principles of kinetic theory
of gases} and the related construction of Boltzmann kinetic equation \cite%
{Grad}\footnote{%
This was also the subject of his last public lecture before his death
occurred on November 17, 1986. It was delivered as invited opening speech at
the plenary session of the 15th RGD Symposium, held in Grado, Italy in July
of the same year \cite{Grad1986}. Unfortunately immediately after the
presentation he suffered the symptoms of a heart attack from which he never
fully recovered.}. Grad's approach in fact represents a first attempt at an
axiomatic formulation of the microscopic statistical description, \textit{%
i.e.,} based on classical statistical mechanics, for a classical $N-$body
system $S_{N}$ in which all particles have, at \ least in principle, a
finite-size. This is realized via the construction of the $N-$body
probability density functions (PDF) $\rho ^{(N)}(\mathbf{x},t)$ for a closed
classical $N-$body system $S_{N}$, \textit{i.e.,} in which the number of
particles ($N$) remains constant. By assumption $S_{N}$\ is immersed in a
bounded and simply-connected subspace $\Omega $\ of the Euclidean space $%
%TCIMACRO{\U{211d} }%
%BeginExpansion
\mathbb{R}
%EndExpansion
^{3}$\ having rigid boundary\textbf{\ }$\delta \Omega $ and a finite
canonical measure $L^{3}=\mu (\Omega )$ (with $L$ being the corresponding
configuration-space characteristic scale length). In particular, $S_{N}$ is
identified, as in the case of Boltzmann kinetic theory, with the ensemble of
$N$ like smooth hard spheres of diameter $\sigma $ and mass $m,$ each one
being labelled by its Newtonian state $\mathbf{x}_{j}\equiv \left\{ \mathbf{r%
}_{j},\mathbf{v}_{j}\right\} $ (with $\mathbf{r}_{j}$ and $\mathbf{v}_{j}$,
for all $j=1,N$, denoting the particle center-of-mass position and velocity)
and $\mathbf{x}\equiv \left\{ \mathbf{x}_{1},..,\mathbf{x}_{N}\right\} $
denoting the state of $S_{N}$ spanning the corresponding phase space $\Gamma
^{N}\equiv \left( \Gamma _{1}\right) ^{N}$. By assumption $\mathbf{x}$
evolves in time from an arbitrary initial state $\mathbf{x}(t_{o})\equiv
\mathbf{x}_{o}$ due to instantaneous particle collisions occurring at
discrete collision times $\left\{ t_{i}\right\} \equiv \left\{ t_{i},i\in
%TCIMACRO{\U{2115} }%
%BeginExpansion
\mathbb{N}
%EndExpansion
\right\} $. The collisions themselves are realized either by means of unary,
binary or - in principle - arbitrary multiple elastic collisions occurring
among the particles of $S_{N}$ and/or with its rigid boundary $\delta \Omega
$, the latter being assumed stationary with respect to a suitable inertial
frame (see Figure 1). Accordingly, for a collision event at time $t_{i}$
involving $k$ particles (for $k=1,..,N$) this means that the corresponding
\emph{incoming} and \emph{outgoing} states (\textit{i.e.,} occurring
immediately before and after collision),\ namely the lower and upper limits
(for $j=1,..,k$) $\lim_{t\rightarrow t_{i}^{\pm }}\mathbf{x}_{j}(t)\equiv
\mathbf{x}_{j}^{(\pm )}(t_{i})\equiv \left\{ \mathbf{r}_{j}(t_{i}),\mathbf{v}%
_{j}^{(\pm )}(t_{i})\right\} $ are related by the so-called \emph{elastic
collision laws}. Thus, for example, a binary collision event $(1,2)$ between
particles$\ 1$ and $2$ occurs at time $t=t_{i}$ provided: 1) at the
collision time $t_{i}$ the same particles are in instantaneous mutual
contact so that $\left\vert \mathbf{r}_{2}-\mathbf{r}_{1}\right\vert =\sigma
$; 2) before collision the particles are approaching each other, in the
sense that the relative incoming velocity $\mathbf{v}_{12}^{(-)}\equiv
\mathbf{v}_{1}^{(-)}-\mathbf{v}_{2}^{(-)}$ is such that $\mathbf{n}%
_{12}\cdot \mathbf{v}_{12}^{(-)}<0$, with $\mathbf{n}_{12}$ denoting the
unit vector $\mathbf{n}_{12}=vers\left\{ \mathbf{r}_{1}-\mathbf{r}%
_{2}\right\} $.

\begin{center}
\begin{figure}[h!]
\includegraphics[height=.30 \textheight]{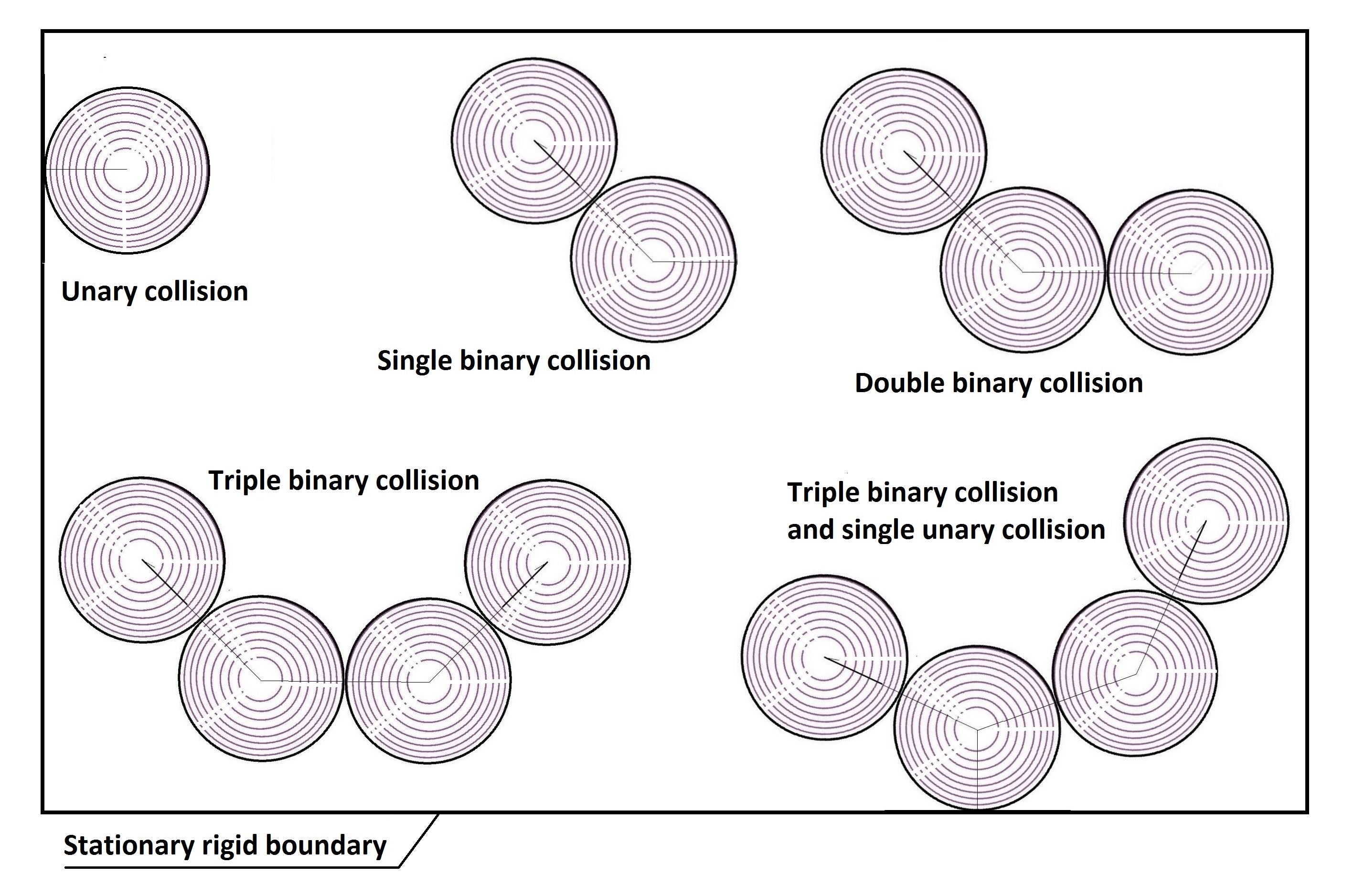}
\caption{Unary, binary and multiple collisions.}
\end{figure}
\end{center}

\noindent The corresponding $2-$particle elastic collision laws are then
realized by the velocity transformations indicated in Figure 2. Similarly,
the occurrence at time $t=t_{i}$ of a double binary collision event $%
(1,2)-(2,3)$ requires simultaneously 1) that $\left\vert \mathbf{r}_{2}-%
\mathbf{r}_{1}\right\vert =\sigma $ and $\left\vert \mathbf{r}_{2}-\mathbf{r}%
_{3}\right\vert =\sigma $ with\ $\left\vert \mathbf{r}_{1}-\mathbf{r}%
_{3}\right\vert >\sigma $, and in addition 2) that $\mathbf{n}_{12}\cdot
\mathbf{v}_{12}^{(-)}<0$ and $\mathbf{n}_{23}\cdot \mathbf{v}_{23}^{(-)}<0,$
namely that particles $1$ and $2$ as well as $2$ and $3$ are both
approaching each other. Analogous collision laws can be established for
arbitrary higher-order multiple collisions. As a consequence, the
time-evolution of the $N-$body state $\mathbf{x}\left( t\right) \equiv
\left\{ \mathbf{x}_{1}\left( t\right) ,..,\mathbf{x}_{N}\left( t\right)
\right\} $ remains by construction uniquely prescribed for all $t\in I\equiv
\mathbb{R} $\footnote{%
this identifies the so-called \emph{Boltzmann-Sinai\ classical dynamical
system}.}.

\begin{center}
\begin{figure}[h!]
\includegraphics[height=.25 \textheight]{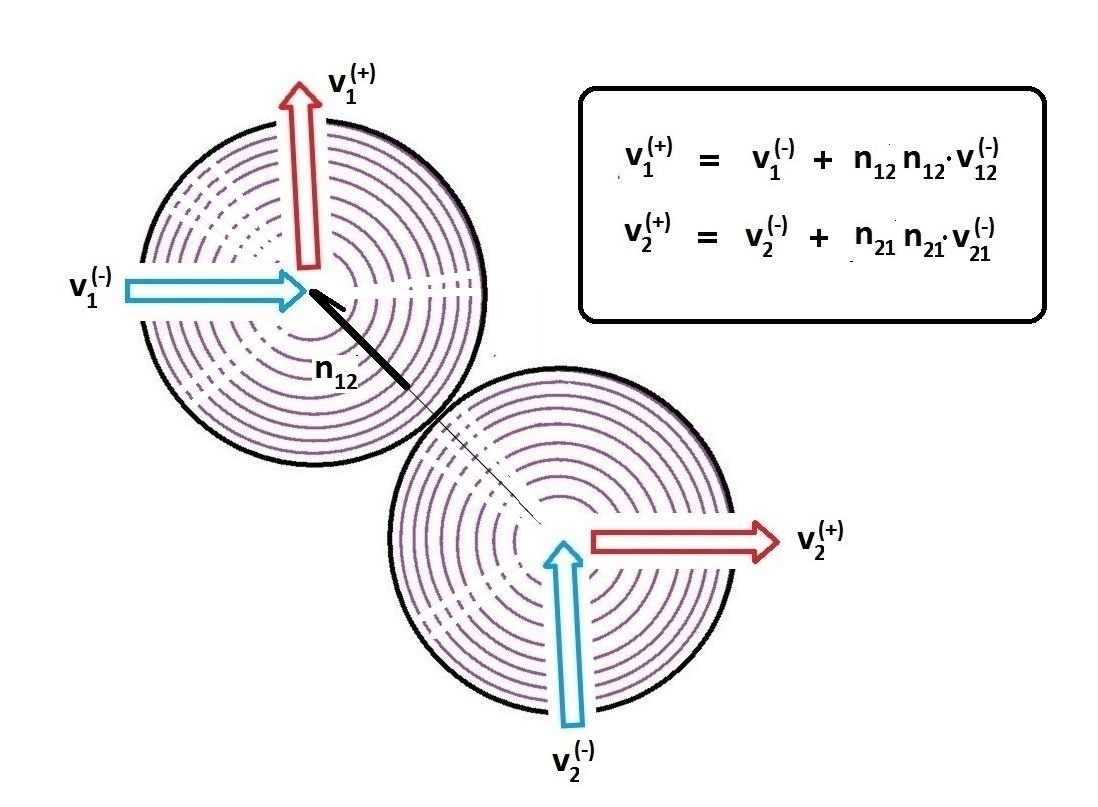}
\caption{Elastic binary collision law.}
\end{figure}
\end{center}

Grad's approach became popular in the subsequent literature being adopted by
most authors (see for example Cercignani \cite{Cercignani1975}). In
particular, it was instrumental to overcome the notorious Loschmidt paradox
\cite{Loschmidt}, \textit{i.e.,} the claim that Boltzmann H-theorem might
conflict with the time-reversal invariance of the Boltzmann-Sinai CDS. The
subsequent response given by Boltzmann \cite{Boltzmann1896c} was in itself
manifestly self-contradictory since he conceded that "\emph{H-theorem could
be violated in some cases}" \footnote{%
according to Drory \cite{Droty2008} this might be the reason of the
Boltzmann's dramatic depression that led him to commit suicide on Sept.5,
1906 during his family Summer vacation in the castle of Duino, near Trieste,
Italy.}. This circumstance might/should occur, according to Boltzmann, for
suitable initial conditions of the $1-$body PDF denoted as "high-entropy"
states (see related discussion in Ref. \cite{Lebowitz1993}). Such a view,
however, based on the proof given by Cercignani and Lampis in Ref. \cite%
{Crcignani e Lampis}, appears incorrect. In fact, contrary to Loschmidt
objection, the Boltzmann-H theorem preserves its full validity under
time-reversal, being - in actual fact - time-reversal invariant. The
property arises because, once the time-reversal transformation\ is
performed, the appropriate corresponding causal prescription must be adopted
at the kinetic level for the time evolution of the $2-$body PDF occurring
across arbitrary binary collision events (see next section below). Such a
prescription affects in turn the realization of the Boltzmann collision
operator and of the Boltzmann kinetic equation too. As a consequence, in
contrast to the aforementioned Boltzmann interpretation, the inequality
characterizing the Boltzmann-H theorem remains necessarily unchanged under
the action of the same transformation.

Despite this conclusion, the fundamental property of \emph{decay to kinetic
equilibrium} (DKE) (also usually referred to as \emph{Macroscopic
irreversibility}) which is implied by the Boltzmann H-theorem remains true.
Accordingly, for suitably-smooth initial conditions the solution of the
Boltzmann kinetic equation $\rho _{1}(\mathbf{x}_{1},t)$ in the limit $%
t\rightarrow +\infty $ should decay to a spatially-homogeneous Maxwellian
PDF of the type%
\begin{equation}
\rho _{M}(\mathbf{v}_{1})=\frac{n_{o}}{\pi ^{3/2}\left( 2T_{o}/m\right)
^{3/2}}\exp \left\{ -\frac{m\left( \mathbf{v}_{1}-\mathbf{V}_{o}\right) ^{2}%
}{2T_{o}}\right\} ,  \label{P1-3}
\end{equation}%
with $\left\{ n_{o}>0,T_{o}>0,\mathbf{V}_{o}\right\} $\emph{\ }being
suitable constant fluid fields. Such a result is highly non-trivial because
it should rely on a global existence theorem for the Boltzmann kinetic
equation. However, according to Villani\emph{\ }\cite{Villani}, "\emph{...\
present-day mathematics is unable to prove }(such a result) \emph{rigorously
and in satisfactory generality."} the obstacle being that it is not known%
\emph{\ "..whether solutions of the Boltzmann equation are smooth enough,
except in certain \ particular cases".}{\small \ } Until recently \cite{noi7}
the answer to this question, clearly of fundamental importance also for the
practical applications of the Boltzmann equation, has remained elusive.

In addition, certain intrinsically physics-related aspects of Grad's
approach, as well as, incidentally, of the same one originally developed by
Boltzmann, were similarly left unsolved, thus actually preventing its
straightforward extension to the treatment of dense and/or granular fluids.
In fact, both theories are actually specialized to the treatment of the
so-called \emph{Boltzmann-Grad limit}. For definiteness let us denote by $%
\Delta \equiv \frac{4\pi N\sigma ^{3}}{3L^{3}}$ the \emph{global diluteness
parameter}. Then the prescription of the Boltzmann-Grad limit involves,
besides suitable smoothness conditions (see discussion below), invoking in
addition: A) first, the validity of the so-called\emph{\ dilute-gas
asymptotic ordering} for $N,\sigma $ and $L$, which is obtained by invoking
the asymptotic condition $N\equiv \frac{1}{\varepsilon }\gg 1,$ together
with the requirements that $\sigma $ and the scale length $L$ be ordered
respectively so that $\sigma \sim O(\varepsilon ^{1/2})$ and $L\sim
O(\varepsilon ^{0}).$ Then the second ordering implies in turn that $\Delta $
must be considered of order $O(\varepsilon ^{1/2})$, \textit{i.e.}, as
corresponds to a globally dilute system; B) second, \emph{the continuum limit%
}, obtained letting $\varepsilon \equiv \frac{1}{N}\rightarrow 0^{+}$.
Therefore, the issue arises of the proper extension of Boltzmann's and
Grad's kinetic theories to the statistical treatment of: 1) \emph{granular
systems},\ i.e,, in which constituent particles have\textit{\ }a finite-size
\cite{Enskog}; 2) \emph{finite systems}, \textit{i.e.,} statistical
ensembles having a finite number ($N$) of particles; 3) emph{dense} or \emph{%
locally dense systems}, \textit{i.e.,} for which the $\Delta \sim O(1)$ or
the characteristic scale length of $1-$body PDF becomes comparable with the
size of the particles $\sigma $. A critical question in this connection is
the physical basis of the involved microscopic statistical description \cite%
{Grad,noi1} adopted by Grad. For this purpose it is useful to briefly
analyze the basic assumptions laying at the basis of his approach.

\section{Grad's heritage}

The axiomatic approach developed by Grad in his 1958 paper consists actually
in two distinct steps. The first one is realized by the \emph{global unique}
\emph{prescription} of the $N-$body probability density function (PDF) $\rho
^{(N)}(\mathbf{x},t)$, \textit{i.e.,} holding identically in the extended $%
N- $body phase space $\Gamma ^{N}\times I.$ The second one, by the
construction of the associated BBGKY hierarchy of equations obtained for all
$s=1,..,N-1$ in terms of the reduced $s-$body PDF's $\rho _{s}^{(N)}(\mathbf{%
x}_{1},..,\mathbf{x}_{s},t)$. Regarding the first topic, once the initial
condition $\rho ^{(N)}(\mathbf{x,}t_{o})=\rho _{o}^{(N)}(\mathbf{x})$ is set
- with $\rho _{o}^{(N)}(\mathbf{x})$ denoting an initial PDF belonging to a
suitable functional class $\left\{ \rho _{o}^{(N)}(\mathbf{x})\right\} $ -
the task involves fulfilling the following two basic requirements: 1) \emph{%
Physical prerequisite \#1:} the first consists in the realization of the
functional setting of $\rho ^{(N)}(\mathbf{x},t)$, namely of the functional
class $\left\{ \rho ^{(N)}(\mathbf{x},t)\right\} $ and hence also the
corresponding one for the initial condition, namely $\left\{ \rho _{o}^{(N)}(%
\mathbf{x})\right\} $. 2) \emph{Physical prerequisite \#2 :} the second one
is the prescription of the \emph{collision boundary condition} (CBC)\textit{%
, i.e.,} the relationship between the \emph{incoming} and \emph{outgoing} $%
N- $body PDF's holding at an arbitrary collision time $t_{i}\in \left\{
t_{i}\right\} ,$ namely $\lim_{t\rightarrow t_{i}^{\pm }}\rho ^{(N)}(\mathbf{%
x}(t),t)\equiv \rho ^{(N)(\pm )}(\mathbf{x}^{(\pm )}(t_{i}),t_{i}),$ where
in the case $\left( -\right) $ the assumption of left-continuity is
introduced requiring\textbf{\ }$\rho ^{(N)}(\mathbf{x}^{(-)}(t_{i}),t_{i})%
\equiv \rho ^{(N)(-)}(\mathbf{x}^{(-)}(t_{i}),t_{i}).$ Such a relationship
must be prescribed in such a way to permit one to represent uniquely either $%
\rho ^{(N)(+)}(\mathbf{x}^{(+)}(t_{i}),t_{i})$ in terms of $\rho ^{(N)}(%
\mathbf{x}^{(-)}(t_{i}),t_{i})$, yielding in this way the \emph{causal CBC},
or \textit{viceversa} $\rho ^{(N)}(\mathbf{x}^{(-)}(t_{i}),t_{i})$ in terms
of $\rho ^{(N)(+)}(\mathbf{x}^{(+)}(t_{i}),t_{i})$ (\emph{anti-causal CBC}).
In both cases it is obvious that the appropriate prescriptions should be
determined uniquely based on the axioms of classical statistical mechanics.
It is interesting, in this regard, to point out the choices adopted by Grad.
Regarding the first one, $\left\{ \rho ^{(N)}(\mathbf{x},t)\right\} $ was
identified with the class of stochastic PDF's, \textit{i.e.,} represented by
smooth ordinary functions. This allowed him to discover that the BBGKY
hierarchy depends functionally on binary collisions only, because multiple
collisions involve surface integrals on phase-space subset of lower
dimension (see related discussion in Ref.\cite{Cercignani1975}). However,
the realization of the BBGKY hierarchy depends on the specific prescription
adopted for the CBC. For this purpose he adopted the same choice originally
introduced by Boltzmann \cite{Boltzmann1972} in his construction of the
Boltzmann equation.\textbf{\ }This follows by requiring that the\ $N-$body
PDF $\rho ^{\left( N\right) }(\mathbf{x},t)$ should remain constant across
arbitrary collisions, \textit{i.e.,} requiring for arbitrary collision times
$t_{i}$ the so-called \emph{PDF-conserving CBC }Lagrangian conservation law%
\begin{equation}
\rho ^{(N)(+)}(\mathbf{x}^{\left( +\right) }(t_{i}),t_{i})=\rho ^{\left(
N\right) }(\mathbf{x}^{\left( -\right) }(t_{i}),t_{i}).
\label{PDF-conserving-CBC}
\end{equation}%
Hence, validity of the causality principle requires suitably representing
the surface integrals appearing in the BBGKY hierarchy so that $\rho
^{\left( N\right) (+)}(\mathbf{x}^{\left( +\right) }(t_{i}),t_{i})$ should
be represented in terms of $\rho ^{\left( N\right) }(\mathbf{x}^{\left(
-\right) }(t_{i}),t_{i})$ and not vice-versa \footnote{%
their exchange implies in fact, as pointed out in Ref. \cite{Crcignani e
Lampis}, a change of signature in the entropy production rate.}. The same
equation (\ref{PDF-conserving-CBC}) can formally be written also in the
equivalent Eulerian form. This requires for arbitrary $t\in I$ the equation
\begin{equation}
\rho ^{\left( N\right) (+)}(\mathbf{x}^{\left( +\right) },t)=\rho ^{\left(
N\right) }(\mathbf{x}^{(-)},t)  \label{EULERIANPDF-conserving CBC}
\end{equation}%
to hold, with $\mathbf{x}^{(-)}$ and $\mathbf{x}^{\left( +\right) }=\mathbf{x%
}^{\left( +\right) }(\mathbf{x}^{(-)})$ denoting colliding states,
respectively an arbitrary incoming and the corresponding outgoing one
prescribed by means of the elastic collision laws. One can readily find out
the key assumption underlying Grad's choice (\ref{PDF-conserving-CBC}).
Indeed, in any collisionless time interval $I_{i}\equiv \left] t_{i},t_{i+1}%
\right[ $ between two consecutive collision times $t_{i}$ and $t_{i+1}$, the
same PDF must manifestly satisfy the integral Liouville equation
\begin{equation}
\rho ^{\left( N\right) }(\mathbf{x}(t+\tau ),t+\tau )=\rho ^{\left( N\right)
}(\mathbf{x}\left( t\right) ,t),  \label{integral Liouville}
\end{equation}%
for all $t+\tau $\emph{\ }and\emph{\ }$t$, belonging to a given
collisionless time interval $I_{i}$. It follows that Eq.(\ref%
{PDF-conserving-CBC}) is therefore equivalent to the requirement that the $%
N- $body PDF should be \emph{globally conserved} along an arbitrary
Lagrangian trajectory $\left\{ \mathbf{x}\left( t\right) \right\} $.

\section{The new "\textit{ab initio}" axiomatic approach}

In a series of recent papers \cite{noi1,noi2,noi3,noi4,noi5,noi6,noi7} a new
solution has been adopted for \emph{Physical prerequisites \#1 and 2,}
referred to as "\textit{ab initio}" \emph{approach }to the microscopic
statistical description of\emph{\ }$S_{N}$. This is based on a careful
rethinking of Grad 1958 axiomatic approach involving in place of his choices
for the same prerequisites, respectively the introduction of suitable \emph{%
extended functional setting} for the $N-$body PDF, \textit{i.e.,} an
appropriate prescription of $\left\{ \rho ^{(N)}(\mathbf{x},t)\right\} $,
and \emph{modified collision boundary condition} (MCBC) to hold at arbitrary
collision events. To start with one notices the peculiar non-local feature
of the Lagrangian or Eulerian CBC indicated above (see Eqs.(\ref%
{PDF-conserving-CBC}) and (\ref{EULERIANPDF-conserving CBC})), which relates
the incoming and outgoing PDF's evaluated at different phase-space states.
The question which arises is whether a \emph{locality} \emph{prescription}
for the appropriate collision boundary conditions should, instead, be
adopted. More precisely, this means that CBC should be realized by means of
a \emph{local relationship} between $\rho ^{(N)(+)}$ and $\rho ^{(N)(-)}$
when both are evaluated \emph{at the same state}. In other words, this
requires prescribing the functional form of the outgoing PDF in terms of the
same outgoing state only,\textit{\ i.e.}, after collision.

An additional feature emerges by inspection of Grad's approach. In fact,
although Eq.(\ref{PDF-conserving-CBC}) can just be viewed as a restatement
of the Liouville equation valid across collision times, the validity of Eq.(%
\ref{PDF-conserving-CBC}) is actually non-mandatory.\ Indeed in order to
satisfy the axiom of probability conservation it suffices that the integral
Liouville equation holds only in the sense indicated above by Eq.(\ref%
{integral Liouville}), \textit{i.e.,} when $t$ and $t+\tau $ belong to the
same collisionless time subset $I_{i}$. In order to clarify this point let
us notice in fact that, based on the axioms of classical statical mechanics,
the deterministic $N-$body PDF must be necessarily an admissible particular
solution of the Liouville equation \cite{noi2}. This means that the
functional class $\left\{ \rho ^{(N)}(\mathbf{x},t)\right\} $ should
include, besides ordinary functions, also distributions and in particular
the \emph{deterministic Dirac} \emph{delta} $N-$body PDF $\rho _{H}^{(N)}(%
\mathbf{x},t)\equiv \delta (\mathbf{x}-\mathbf{x}(t))$. The
physically-consistent characterization of the collision boundary conditions
should therefore permit the treatment of such a case. However, by
construction it follows that same PDF must satisfy the collision boundary
condition requiring simultaneously that%
\begin{equation}
\left\{
\begin{array}{c}
\rho _{H}^{(N)}(\mathbf{x},t_{i})\equiv \rho _{H}^{(N)(-)}(\mathbf{x}%
,t_{i})=\delta (\mathbf{x}-\mathbf{x}^{(-)}(t_{i})), \\
\rho _{H}^{(N)(+)}(\mathbf{x},t_{i})\equiv \delta (\mathbf{x}-\mathbf{x}%
^{(+)}(t_{i})).%
\end{array}%
\right.  \label{MCBC-0}
\end{equation}%
On the other hand, an arbitrary stochastic PDF $\rho ^{\left( N\right) }(%
\mathbf{x}(t),t)$\ can always be represented in terms of the convolution
integral $\rho ^{\left( N\right) }(\mathbf{x}(t),t)=\int d\mathbf{x}\rho
^{\left( N\right) }(\mathbf{x},t)\delta (\mathbf{x}-\mathbf{x}(t))$, which
means that\textbf{\ }$\rho ^{(+)\left( N\right) }(\mathbf{x}^{\left(
+\right) }(t_{i}),t_{i})=\int d\mathbf{x}\rho ^{\left( N\right) }(\mathbf{x}%
,t_{i})\delta (\mathbf{x}-\mathbf{x}^{(+)}(t_{i}))$ while $\rho ^{(-)\left(
N\right) }(\mathbf{x}^{\left( -\right) }(t_{i}),t_{i})=\int d\mathbf{x}\rho
^{\left( N\right) }(\mathbf{x},t_{i})\delta (\mathbf{x}-\mathbf{x}%
^{(-)}(t_{i}))$ so that in particular $\rho ^{(-)\left( N\right) }(\mathbf{x}%
^{\left( -\right) }(t_{i}),t_{i})\equiv \rho ^{\left( N\right) }(\mathbf{x}%
^{\left( -\right) }(t_{i}),t_{i})$. Hence the correct realization of the CBC
for stochastic PDF's is necessarily given by the causal relationship \cite%
{noi2}%
\begin{equation}
\rho ^{(+)\left( N\right) }(\mathbf{x}^{\left( +\right) }(t_{i}),t_{i})=\rho
^{\left( N\right) }(\mathbf{x}^{(+)}(t_{i}),t_{i}),  \label{MCBC}
\end{equation}%
to be referred to as modified CBC (MCBC) in Lagrangian form. The
corresponding Eulerian condition holding for arbitrary $(\mathbf{x}^{\left(
+\right) },t)$ is therefore provided by $\rho ^{(+)\left( N\right) }(\mathbf{%
x}^{\left( +\right) },t)=\rho ^{\left( N\right) }(\mathbf{x}^{(+)},t)$. The
physical interpretation of Eq.(\ref{MCBC}) is intuitive. It can be viewed,
in fact, as the jump condition for the $N-$body PDF along the phase-space
Lagrangian trajectory $\left\{ \mathbf{x}(t)\right\} $ for an ensemble of $N$
tracer particles \cite{noi0} following the same deterministic trajectory and
undergoing a collision event at time $t_{i}.$ For these particles the same $%
N-$body PDF $\rho ^{\left( N\right) }(\mathbf{x},t)$ must obviously be \emph{%
considered as prescribed }and therefore it is manifest that its form \emph{%
cannot be affected }by the said collision event occurring for the test
particles. Notice, additionally, that the $N-$body Dirac delta itself can be
considered as the limit of the sequence\textbf{\ }$\left\{ \rho
_{H(i)}^{(N)}(\mathbf{x},\mathbf{x}(t)),i\in
%TCIMACRO{\U{2115} }%
%BeginExpansion
\mathbb{N}
%EndExpansion
\right\} $\textbf{\ }in which each function\textbf{\ }$\rho _{H(i)}^{(N)}(%
\mathbf{x},\mathbf{x}(t))$\textbf{\ }is a\textbf{\ }PDF satisfying MCBC \cite%
{noi2}.\textbf{\ }Hence Eq.(\ref{MCBC}) is a direct consequence of Eqs. (\ref%
{MCBC-0}) which holds in the case of arbitrary $N-$body PDF's different from
the deterministic one \cite{noi2}. We notice that in Eq.(\ref{MCBC}) both
the finite size of the particles and the elastic collision laws are
explicitly taken into account. Such a choice can be shown to be of critical
importance for the statistical treatment of granular or dense gases in which
the finite size of the hard spheres becomes relevant \cite{noi3}. In
addition, one can show that MCBC warrants the conservation laws of the
corresponding collision operators appearing in the BBGKY hierarchy \cite%
{noi4} and the existence of the customary Boltzmann collision invariants
\cite{noi5}.

Finally, it is obvious that MCBC must apply also to the Boltzmann equation.
Nevertheless, as shown in Ref. \cite{noi1} (see also Refs.\cite{noi3,noi7}),
provided the $1-$body PDF is sufficiently smooth the distinction between
MCBC and PDF-conserving CBC becomes effectively irrelevant in such a case.

\section{Physical implications}

The features outlined above imply a radical conceptual change of viewpoint
in kinetic theory which sets it apart from the Boltzmann and Grad
statistical theories as well as Enskog approach to finite-size hard-sphere
systems \cite{Enskog} (see also related discussion in Ref. \cite{noi3}). The
consequences of the new Physical prerequisites are, in fact, far-reaching.
Indeed, as shown in Ref. \cite{noi3}, the "\textit{ab initio}" approach
leads to the establishment of a kinetic equation, realized by the \emph{%
Master kinetic equation. }The Master kinetic equation for the corresponding
stochastic reduced $1-$body PDF can be represented in terms of the
integro-differential equation%
\begin{equation}
\left[ \frac{\partial }{\partial t}+\mathbf{v}_{1}\cdot \frac{\partial }{%
\partial \mathbf{r}_{1}}\right] \rho _{1}^{(N)}(\mathbf{x}_{1},t)=\mathcal{C}%
_{1}\left( \rho _{1}^{(N)}|\rho _{1}^{(N)}\right) ,  \label{MASTER EQUATION}
\end{equation}%
where the operators $C_{1}\left( \rho _{1}^{(N)}|\rho _{1}^{(N)}\right) $\
identifies the Master collision operator. Consistent with the causality
principle, MCBC as well the existence of the customary Boltzmann collision
invariants \cite{noi5} this is found to be%
\begin{gather}
\mathcal{C}_{1}\left( \rho _{1}^{(N)}|\rho _{1}^{(N)}\right) \equiv
K_{n}\int\limits_{U_{1}}d\mathbf{v}_{2}\int^{\mathbf{(-)}}d\Sigma _{21}
\label{M-O} \\
\left\vert \mathbf{v}_{21}\cdot \mathbf{n}_{21}\right\vert \overline{\Theta }%
^{\ast }(\mathbf{x}_{2})\left[ \rho _{2}^{(N)}(\mathbf{x}_{1}^{(+)},\mathbf{x%
}_{2}^{(+)},t)-\rho _{2}^{(N)}(\mathbf{x}_{1},\mathbf{x}_{2},t)\right] .
\notag
\end{gather}%
Here the notation is given in accordance with Ref.\cite{noi3}. Thus $%
K_{n}\equiv \left( N-1\right) \sigma ^{2}$ is the Knudsen coefficient, $%
U_{1}\equiv
%TCIMACRO{\U{211d} }%
%BeginExpansion
\mathbb{R}
%EndExpansion
^{3}$ is the $1-$body velocity space, $\rho _{2}^{(N)}(\mathbf{x}_{1},%
\mathbf{x}_{2},t)\equiv \rho _{1}^{(N)}(\mathbf{r}_{1},\mathbf{v}_{1},t)\rho
_{1}^{(N)}(\mathbf{r}_{2},\mathbf{v}_{2},t)\frac{k_{2}^{(N)}(\mathbf{r}_{1},%
\mathbf{r}_{2},t)}{k_{1}^{(N)}(\mathbf{r}_{1},t)k_{1}^{(N)}(\mathbf{r}_{2},t)%
}$ the $2-$body PDF, the symbol $\int^{(-)}d\Sigma _{21}$ denotes
integration on the subset of the solid angle of incoming particles for which
$\mathbf{v}_{12}\cdot \mathbf{n}_{12}<0,$ while in the integrand $\mathbf{r}%
_{2}=\mathbf{r}_{1}+\sigma \mathbf{n}_{21}$ labels the position of particle "%
$2$" which is colliding and hence is in contact with particle "$1$"$.$
Furthermore, the position-dependent functions $k_{1}^{(N)}(\mathbf{r}_{1},t)$
and $k_{2}^{(N)}(\mathbf{r}_{1},\mathbf{r}_{2},t)$ identify so-called
\textbf{\ }$1-$ and $2-$body occupation coefficients. Remarkably, in
difference with Enskog's kinetic equation \cite{Enskog}, in the context of
the "\textit{ab initio}" approach they are uniquely prescribed \cite{noi3}.
From the physical point of view, these coefficients are related to the
occupation domain of the hard spheres in the configuration space $\Omega $
arising due to their finite size. Finally $\overline{\Theta }^{\ast }(%
\mathbf{x}_{2})$ denotes the domain theta function $\overline{\Theta }^{\ast
}(\mathbf{x}_{2})\equiv \overline{\Theta }\left( \left\vert \mathbf{r}_{2}-%
\frac{\sigma }{2}\mathbf{n}_{2}\right\vert -\frac{\sigma }{2}\right) $, with
$\mathbf{n}_{2}$ denoting the inward normal unit vector to the boundary $%
\partial \Omega $\ and $\overline{\Theta }$ being the strong theta function.
The really remarkable feature of Eq.(\ref{MASTER EQUATION}) is that it is
exact, \textit{i.e.,} it holds for an $N-$body hard-sphere system having an
arbitrary \emph{finite number} of particles and for hard spheres having
\emph{finite-size}, namely with finite diameter $\sigma >0,$ and \emph{%
finite mass} $m$\ too.\textbf{\ }This follows as a consequence of both the
extended functional setting indicated above and the MCBC adopted for the
same $N-$body PDF\emph{.}

Based on the construction of the Master kinetic equation, a host of new
exciting developments in kinetic theory have opened up. In fact, in several
respects the new "\textit{ab initio}" approach differs significantly from
previous literature. The main difference arises because of the
non-asymptotic character of the new kinetic equation, i.e., the fact that it
applies to arbitrary smooth hard sphere systems for which the finite number
and size of the constituent particles is accounted for. Possible
applications are ubiquitous (see Figure 3). These include examples such as:
1) \emph{Example \#1:} \emph{Environmental} and \emph{material-science
granular fluids} (ambient atmosphere, sea-water and ocean dynamics, etc.);
2) \emph{Example \#2:} \emph{Biological granular fluids} (bacterial motion
in fluids, cell-blood dynamics in the human body, blood-vessels,
capillaries, etc.); 3) \emph{Example \#3:} \emph{Industrial granular fluids}
(grain or pellet dynamics in metallurgical and chemical processes, air and
water pollution dynamics, etc.); 4) \emph{Example \#4:} \emph{Geological
fluids} (slow dynamics of highly viscous granular fluids, inner Earth-core
dynamics, etc.).

\begin{center}
\begin{figure}[h!]
\includegraphics[height=.25 \textheight]{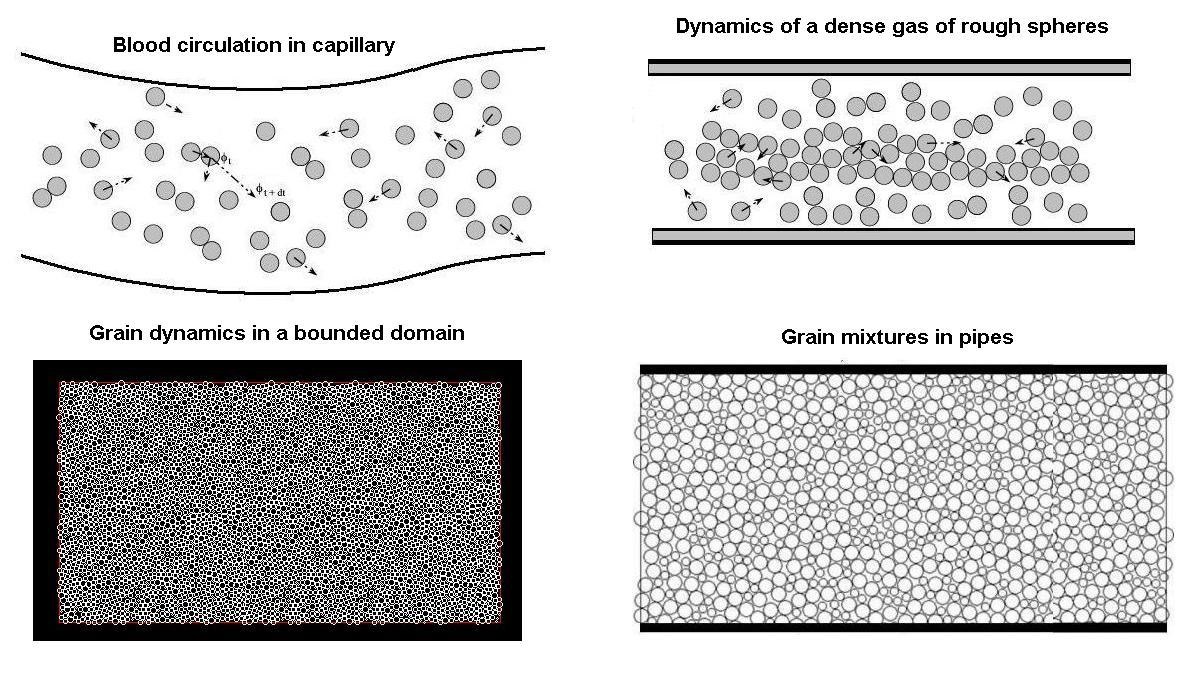}
\caption{Examples of dense granular flows}
\end{figure}
\end{center}

In most of the cases indicated above a self-consistent statistical
description to be based on classical statistical mechanics, not mentioning a
kinetic equation, was previously missing or largely unsatisfactory. The "%
\textit{ab initio}" statistical approach provides such a missing link which
is based on the Master kinetic equation established in Ref. \cite{noi3}.

Nevertheless in validity of the Boltzmann-Grad limit the Master equation
reduces exactly to the Boltzmann kinetic equation \cite{noi3,noi7}. In fact,
one notices that if such a limit is introduced the limit $(N-1)\sigma
^{2}\rightarrow N\sigma ^{2}$ applies and an arbitrary stochastic and $1-$%
body PDF $\rho _{1}^{(N)}(\mathbf{x}_{1},t)$ can be replaced by its
asymptotic approximation represented by the Boltzmann-Grad limit function $%
\rho _{1}(\mathbf{x}_{1},t).$ In the Master collision operator this occurs,
however, only provided suitable smoothness assumptions hold for the $1-$body
PDF's and\ the $1-$ and $2-$body occupation coefficients $k_{1}^{(N)}(%
\mathbf{r}_{1},t)$ and $k_{2}^{(N)}(\mathbf{r}_{1},\mathbf{r}_{2},t)$. These
require, more precisely that, when the diameter of the particles $\sigma $
becomes infinitesimal, it should be possible to replace a) the $1-$body PDFs
$\rho _{1}^{(N)}(\mathbf{r}_{2},\mathbf{v}_{2}^{(+)},t)$ and $\rho
_{1}^{(N)}(\mathbf{r}_{2},\mathbf{v}_{2},t)$ with their limits $\rho _{1}(%
\mathbf{r}_{1},\mathbf{v}_{2}^{(+)},t)$ and $\rho _{1}(\mathbf{r}_{1},%
\mathbf{v}_{2},t)$ respectively; b) the $1-$ and $2-$body occupation
coefficients $k_{1}^{(N)}(\mathbf{r}_{1},t)$ and $k_{2}^{(N)}(\mathbf{r}_{1},%
\mathbf{r}_{2},t)$ with their limit functions, namely respectively $k_{1}(%
\mathbf{r}_{1},t)=k_{2}(\mathbf{r}_{1},\mathbf{r}_{2},t)=1$ \cite{noi3}.
Moreover, for the same reasons indicated above,\ in the Master collision
operator the integration on the sub-domain $\mathbf{v}_{12}\cdot \mathbf{n}%
_{12}<0$ can be equivalently exchanged with $\mathbf{v}_{12}\cdot \mathbf{n}%
_{12}>0$ while the domain theta function $\overline{\Theta }^{\ast }(\mathbf{%
x}_{2})$ becomes $\overline{\Theta }^{\ast }(\mathbf{x}_{2})\equiv \overline{%
\Theta }\left( \left\vert \mathbf{r}_{2}\right\vert \right) $ so that its
contribution to the collision integral becomes ignorable in the case in
which $\rho _{1}(\mathbf{r}_{2},\mathbf{v}_{2},t)$ is stochastic. Based on
these premises, denoting $\rho _{2}(\mathbf{r}_{1},\mathbf{v}_{1},\mathbf{r}%
_{1},\mathbf{v}_{2},t)\equiv \rho _{1}(\mathbf{r}_{1},\mathbf{v}_{1},t)\rho
_{1}(\mathbf{r}_{1},\mathbf{v}_{2},t)$ the $2-$body PDF, it follows that in
the Boltzmann-Grad limit the Master collision operator becomes%
\begin{eqnarray}
&&\left. \mathcal{C}_{B}\left( \rho _{1}|\rho _{1}\right) \equiv N\sigma
^{2}\int\limits_{U_{1}}d\mathbf{v}_{2}\int^{\mathbf{(+)}}d\Sigma
_{12}\left\vert \mathbf{v}_{12}\cdot \mathbf{n}_{12}\right\vert \right.
\notag \\
&&\left[ \rho _{2}(\mathbf{r}_{1},\mathbf{v}_{1}^{(-)},\mathbf{r}_{1},%
\mathbf{v}_{2}^{(-)},t)-\rho _{2}(\mathbf{r}_{1},\mathbf{v}_{1},\mathbf{r}%
_{1},\mathbf{v}_{2},t)\right] ,  \label{B}
\end{eqnarray}%
which therefore coincides with the customary form of the Boltzmann collision
operator \cite{noi3}. As a consequence it is obvious that also the Boltzmann
H-theorem pointed out originally by Boltzmann himself \cite{Boltzmann1972}
\begin{equation}
\frac{\partial }{\partial t}S(\rho _{1}(t))\geq 0
\label{Boltzmann inequality}
\end{equation}%
necessarily must hold for the limit function $\rho _{1}(\mathbf{x}_{1},t)$.

Despite such a conclusion, the Master kinetic equation has an entirely
different physical character. \ In fact, as shown in Ref. \cite{noi6}, in
sharp contrast with the Boltzmann equation, its solution represented by the $%
1-$body PDF admits an \emph{exact constant H-theorem} in terms of
Boltzmann-Shannon statistical entropy $S(\rho _{1}^{(N)}(t))\equiv
-\int\limits_{\Gamma _{1}}d\mathbf{v}_{1}\rho _{1}^{(N)}(\mathbf{x}%
_{1},t)\ln \rho _{1}^{(N)}(\mathbf{x}_{1},t),$ so that identically
\begin{equation}
\frac{\partial }{\partial t}S(\rho _{1}^{(N)}(t))\equiv 0.
\label{B-.S ENTROPY}
\end{equation}%
\ A number of implications follow concerning the "\textit{ab initio}"
kinetic theory.

The first one is about the long-debated issue about the physical origin of
the Boltzmann entropic inequality (\ref{Boltzmann inequality}) (see Refs.
\cite{Lebowitz1993,noi6, noi8}). This problem can be given a satisfactory
answer within the new kinetic theory. In fact, as shown in Ref. \cite{noi6},
if $\rho _{1}^{(N)}(\mathbf{x}_{1},t)$ is approximated in terms of its
Boltzmann-Grad limit function $\rho _{1}(\mathbf{x}_{1},t)$ the constant
H-theorem (\ref{B-.S ENTROPY})\ is not at variance with the validity of the
same inequality (\ref{Boltzmann inequality}). Indeed, once the replacement $%
\rho _{1}^{(N)}(\mathbf{x}_{1},t)\rightarrow $\ $\rho _{1}(\mathbf{x}_{1},t)$
is made in the corresponding $1-$body Boltzmann-Shannon entropy $S(\rho
_{1}(t))$, this amounts to introduce a related information \textquotedblleft
\textit{error}\textquotedblright . Such an error unavoidably gives rise to
an effective monotonic increase of the Boltzmann--Shannon entropy, so that
the validity of Boltzmann H-theorem can actually be rigorously inferred \cite%
{noi6} for the same class of limit functions.

Second, the same conclusion proves the conjecture proposed originally by
Grad on the physical origin of the Boltzmann entropic inequality. In fact,
in his paper devoted to the principles of kinetic theory \cite{Grad} he
suggested that, from the information-theory viewpoint,\textbf{\ }Boltzmann
H-theorem should be understood in terms of \textquotedblleft information
loss\textquotedblright\ produced by the Boltzmann-Grad limit$.$

Third, the global existence problem for the Master kinetic equation has been
recently addressed based on the corresponding $N-$body Liouville equation
achieved in the context of the new "\textit{ab initio}" approach \cite{noi7}.%
\textbf{\ }In such a case, in fact, global existence and uniqueness for the $%
1-$body PDF $\rho _{1}^{(N)}(\mathbf{x}_{1}\mathbf{,}t)$ can be established
as a consequence of the global unique prescription of the corresponding $N-$%
body PDF $\rho ^{(N)}(\mathbf{x,}t)$ along arbitrary phase-space Lagrangian
trajectories. Indeed, as shown in Ref. \cite{noi3}, in validity of MCBC the
same PDF can always be identified with a suitably-weighted factorizable
solution in terms of the corresponding $1-$body PDF $\rho _{1}^{(N)}(\mathbf{%
x}_{1}\mathbf{,}t)$. As a consequence the same $1-$body PDF $\rho _{1}^{(N)}(%
\mathbf{x}_{1}\mathbf{,}t),$ to be identified in principle with an arbitrary
stochastic PDF, is uniquely globally defined and, thanks again to MCBC, it
necessarily satisfies the Master kinetic equation. These results are
relevant also for the global existence problem posed by Villani \cite%
{Villani}. In fact, analogous conclusions follow in principle also when the
Boltzmann-Grad limit is considered and $\rho _{1}^{(N)}(\mathbf{x}_{1},t)$
is replaced with its limit function $\rho _{1}(\mathbf{x}_{1},t)$. This
requires, however, that the same smoothness conditions needed for the
validity of the Boltzmann collision operator (\ref{B}) should hold globally
in the extended $1-$body phase space $\Gamma _{1}\times I$.

Fourth and last ("\emph{in cauda stat venenum}" paraphrasing Marziale's
famous statement) the question arises of the possible occurrence of the
phenomenon of DKE for an arbitrary stochastic solution of the Master kinetic
equation. The latter refers to the property whereby particular solutions of
the same kinetic equation may decay to a local Maxwellian kinetic
equilibrium of the type (\ref{P1-3}). In the context of Boltzmann and Grad
kinetic theories such a property is customarily associated with the entropic
inequality (\ref{Boltzmann inequality}), implying in turn the occurrence of
DKE for the same PDF's. The question which arises, and remains still to be
answered, is whether, despite the validity of the constant H-theorem (\ref%
{B-.S ENTROPY}), the phenomenon of DKE may arise also for the Master kinetic
equation, i.e., for suitably-smooth particular solutions of the same
equation. The example-case recently pointed out \cite{noi8}, corresponding
to the statistical description of an incompressible Navier-Stokes granular
fluid, suggests that this may be indeed the case. Further investigations in
this direction are under way.

\section{Acknowledgments}

Work developed in part within the research projects: A) the Albert Einstein
Center for Gravitation and Astrophysics, Czech Science Foundation No.
14-37086G; B) the grant No. 02494/2013/RRC \textquotedblleft \textit{kinetick%
\'{y} p\v{r}\'{\i}stup k proud\u{e}n\'{\i} tekutin}\textquotedblright\
(kinetic approach to fluid flow) in the framework of the \textquotedblleft
Research and Development Support in Moravian-Silesian
Region\textquotedblright , Czech Republic: C) the research projects of the
Czech Science Foundation GA\v{C}R grant No. 14-07753P. One of the authors
(M.T.) is grateful to the International Center for Theoretical Physics
(Miramare, Trieste, Italy) for the hospitality during the preparation of the
manuscript.

\end{document}